\documentclass[a4paper,10pt]{article}

\usepackage{graphicx,url}
\usepackage{amsmath,amssymb}

\title{Charm Physics at CDF}
\author{A.~Di~Canto\footnote{Speaker on behalf of the CDF Collaboration.}}
\date{Physikalisches Institut, Ruprecht-Karls-Universit\"{a}t Heidelberg, Germany, and INFN Pisa, Italy.}

\newcommand{\A}{\ensuremath{\mathcal{A}}}
\newcommand{\Acp}{\ensuremath{\A_\textit{CP}}}
\newcommand{\acp}[1]{\ensuremath{\Acp^{\text{#1}}}}
\newcommand{\stat}{\ensuremath{\mathrm{~(stat)}}}
\newcommand{\syst}{\ensuremath{\mathrm{~(syst)}}}
\newcommand{\Dbar}{\ensuremath{\overline{D}{}}}

\begin{document}

\maketitle

\begin{abstract}
The study of the charm quark continues to have wide interest as a possible avenue for the discovery of physics beyond the Standard Model and can as well be used as a tool for understanding the non-perturbative aspects of the strong interactions. Owning to the large production cross-section available at the Tevatron collider and to the flexibility of a trigger for fully hadronic final states, the CDF experiment, in a decade of successful operations, collected millions of charmed mesons decays which can be used to investigate the details of the physics of the production and decay processes of the charm quark. Here we present a brief collection of new CDF results on this subject.
\end{abstract}

\section{Fragmentation of charm quarks}
Heavy quark fragmentation is a non-perturbative process for which Monte Carlo event generators implement only phenomenological models that should be tuned to reproduce the observed properties of hadron production. CDF performs an analysis, described with further details in Ref.~\cite{cdf10704}, that probes the process of quark fragmentation more directly by studying kaons produced during the fragmentation of charm quarks to form a $D_{(s)}^\pm$ meson.

In a data sample corresponding to about $360$\,pb$^{-1}$ of $p\overline{p}$ collisions, we reconstruct about $260\,000$ $D_s^+$ and $140\,000$ $D^+$ mesons decaying to the $\phi(\to K^+K^-)\pi^+$ final state (charge conjugated decays are implied, unless otherwise stated). Promptly produced charmed mesons are statistically separated from products of $b$-hadron decays using the impact parameter distribution of the $D$ candidate. Time-of-flight and ionization energy loss measurements are used to identify kaons and measure their fraction in the sample of maximum-$p_T$ tracks produced in a $\Delta R=\sqrt{\Delta\varphi^2+\Delta\eta^2}\leqslant 0.7$ cone around the $K^+K^-\pi^+$ candidate.  The resulting kaon fractions, separately for same sign and opposite sign categories, are then compared to the predictions of both the string fragmentation model used in \textsc{Pythia} \cite{pythia} and the cluster fragmentation model used in \textsc{Herwig} \cite{herwig}, as a function of several observables: transverse momentum ($p_T$) of the track; invariant mass ($m_{DK}$) of the track (using the kaon mass hypothesis) and the $D$ candidate; difference in rapidity along the fragmentation axis ($\Delta y$) between the track and the $D$ meson. In the opposite sign combination category, where the track in the cone and the $D$ candidate are oppositely charged, we expect the kaon production to be enhanced around $D^+_s$ with respect to $D^+$ mesons since formation of a prompt $D^+_s$ requires conservation of strangeness in the first fragmentation branch. Conversely, in the same sign combination we expect the kaon production to be similar around both $D_s^+$ and $D^+$ mesons since same sign kaons are likely to be produced in later branches of the fragmentation process.

\begin{figure}[t]
\centering
\includegraphics[width=0.8\textwidth]{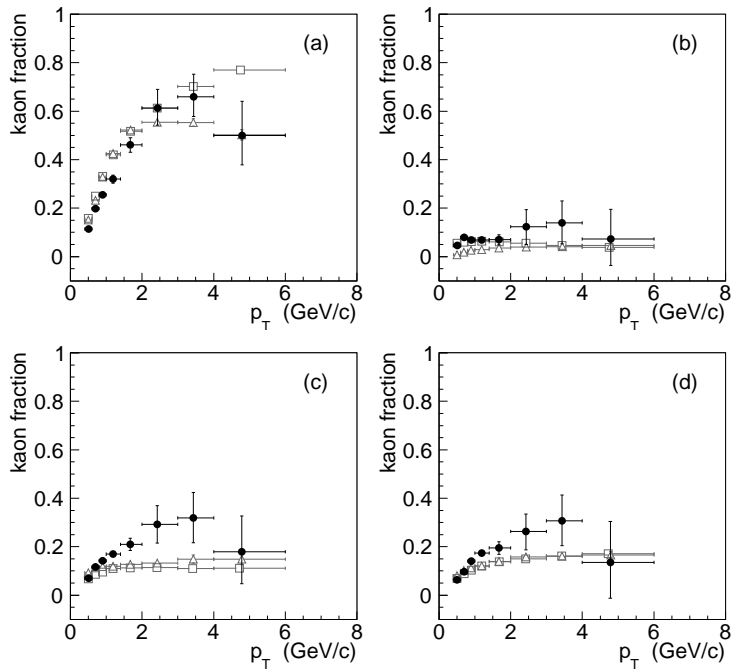}
\caption{Transverse momentum distribution of the measured kaon fraction (solid symbols) in the sample of charged tracks produced in association with $D_s^+$, (a) and (c), and $D^+$ mesons, (b) and (d), separately in the opposite sign, (a) and (b), and same sign, (c) and (d), combinations. Also shown are the kaon fractions calculated using \textsc{Pythia} (open squares) and \textsc{Herwig} (open triangles).}\label{fig01}
\end{figure}

The results of the comparative study show that the $p_T$ distribution for early fragmentation kaons is in better qualitative agreement with predictions than for generic kaons produced in later fragmentation branches, for which the models underestimate the fraction of kaons, as shown in Fig.~\ref{fig01}. Conversely, the $m_{DK}$ and $\Delta y$ distributions indicate that the fragmentation models overestimate the fraction of kaons produced in early stages of the fragmentation process compared to the fraction of generic kaons that are produced in later branches, for which the data show good agreement with predictions.

\section{Search for CP violation in neutral charmed mesons decays}
While CP violation is well established for $B$ and $K$ mesons, this is not the case for charm mesons. First evidence for CP violation in two-body singly-Cabibbo-suppressed $D^0$ decays has been recently reported by the LHCb Collaboration \cite{lhcb}. Whether this is a hint of possible new physics contributions to the decay amplitude or not is not yet clear. It is important to broaden our search for CP violation in further charmed meson decays. Here we present two new searches for CP violation in neutral $D$ mesons decays which are among the world's most sensitive to date.

\subsection{Time-integrated asymmetries in $D^0\to K^0_S\pi^+\pi^-$ decays}
In a data sample corresponding to an integrated luminosity of $6$\,fb$^{-1}$, CDF searches for time-integrated CP asymmetries in the resonant substructure of the three-body $D^0\to K^0_S\pi^+\pi^-$ decay. As the Standard Model expectation of these CP asymmetries is $\mathcal{O}(10^{-6})$, well below the experimental sensitivity, an observation of CP violation would be a clear hint of new physics.

We reconstruct approximately $350\,000$ $D^0\to K^0_S(\to\pi^+\pi^-)\pi^+\pi^-$ candidates with the $D^0$ originating from the strong $D^{*+}\to D^0\pi^+$ decay, in order to unambiguously determine the flavor of the charmed meson at production from the charge of the accompanying pion. Two complementary approaches are used: a full Dalitz fit and a model-independent bin-by-bin comparison of the $D^0$ and $\Dbar^0$ Dalitz plots. We briefly present here only the result of the first approach, a more comprehensive description of the analysis can be found in Ref.~\cite{kspipi}.

\begin{figure}[p]
\centering
\includegraphics[width=0.8\textwidth]{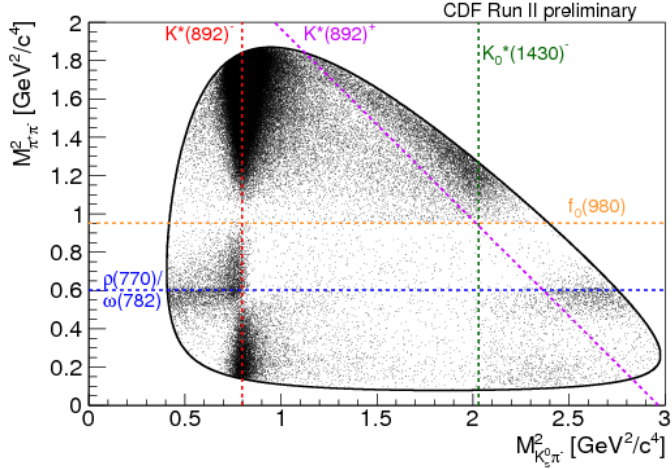}
\caption{Dalitz plot of the reconstructed $D^0\to K_S^0\pi^+\pi^-$ candidates, where some relevant intermediate resonances are indicated by colored dashed lines.}\label{fig04}
\end{figure}

\begin{table}[p]
\centering
\begin{tabular}{lcc}
  \hline
  Resonance & $\mathcal{A}_{\mathrm{FF}}$ (CDF) [\%] & $\mathcal{A}_{\mathrm{FF}}$ (CLEO) [\%]\\
  \hline
  $K^*(892)^-$ & $0.36 \pm 0.33 \pm 0.40$ & $2.5 \pm 1.9\,^{+1.5}_{-0.7}\,^{+2.9}_{-0.3}$\\
  $K_0^*(1430)^-$ & $4.0 \pm 2.4 \pm 3.8$ & $-0.2 \pm 11\,^{+9}_{-5}\,^{+2}_{-1}$\\
  $K_2^*(1430)^-$ & $2.9 \pm 4.0 \pm 4.1$ & $-7 \pm 25\,^{+8}_{-26}\,^{+10}_{-1}$\\
  $K^*(1410)^-$ & $-2.3 \pm 5.7 \pm 6.4$ & $\cdot\cdot\cdot$\\
  $\rho(770)$ & $-0.05 \pm 0.50 \pm 0.08$ & $3.1 \pm 3.8\,^{+2.7}_{-1.8}\,^{+0.4}_{-1.2}$\\
  $\omega(782)$ & $-12.6 \pm 6.0 \pm 2.6$ & $-26 \pm 24\,^{+22}_{-2}\,^{+2}_{-4}$\\
  $f_0(980)$ & $-0.4 \pm 2.2 \pm 1.6$ & $-4.7 \pm 11\,^{+25}_{-7}\,^{+0}_{-5}$\\
  $f_2(1270)$ & $-4.0 \pm 3.4 \pm 3.0$ & $34 \pm 51\,^{+25}_{-71}\,^{+21}_{-34}$\\
  $f_0(1370)$ & $-0.5 \pm 4.6 \pm 7.7$ & $18 \pm 10\,^{+2}_{-21}\,^{+13}_{-6}$\\
  $\rho(1450)$ & $-4.1 \pm 5.2 \pm 8.1$ & $\cdot\cdot\cdot$\\
  $f_0(600)$ & $-2.7 \pm 2.7 \pm 3.6$ & $\cdot\cdot\cdot$\\
  $\sigma_2$ & $-6.8 \pm 7.6 \pm 3.8$ & $\cdot\cdot\cdot$\\
  $K^*(892)^+$ & $1.0 \pm 5.7 \pm 2.1$ & $-21 \pm 42\,^{+17}_{-28}\,^{+22}_{-4}$\\
  $K_0^*(1430)^+$ & $12 \pm 11 \pm 10$ & $\cdot\cdot\cdot$\\
  $K_2^*(1430)^+$ & $-10 \pm 14 \pm 29$ & $\cdot\cdot\cdot$\\
  $K^*(1680)^-$ & $\cdot\cdot\cdot$ & $-36 \pm 19\,^{+9}_{-35}\,^{+5}_{-1}$\\
  \hline
\end{tabular}
\caption{Comparison of the measured fit fraction asymmetries, $\A_{FF}$, for the considered intermediate resonances of the $D^0\to K_S^0\pi^+\pi^-$ decay with the results from the CLEO experiment \cite{cleo}. For the CDF results the first uncertainties are statistical and the second combined systematic. For the CLEO results the first uncertainties are statistical, the second experimental systematic, and the third modeling systematic.}\label{aff}
\end{table}

Fig.~\ref{fig04} shows the Dalitz plot of the reconstructed $D^0\to K_S^0\pi^+\pi^-$ candidates with the most relevant sub-resonant decay modes highlighted. For the first time at a hadron collider, a Dalitz amplitude analysis is applied for the description of the dynamics of the decay. We employ the isobar model and determine the asymmetries between the different $D^0$ and $\Dbar^0$ sub-resonance fit fractions in order to be insensitive to any global instrumental asymmetry in the reconstruction and identification of the candidates of interest. Tab.~\ref{aff} shows the results in comparison with the most recent measurements performed by the CLEO collaboration \cite{cleo}. Our analysis represents a significant improvement in terms of precision, but still no hints of any CP violating effects are found. The measured value for the overall integrated CP asymmetry is
$$\Acp(D^0\to K_S^0\pi^+\pi^-) = \bigl(-0.05\pm 0.57\stat\pm0.54\syst\bigr)\%.$$
Following the procedure described in Ref.~\cite{paper} and assuming no direct CP violation in the $D^0\to K_S^0\pi^+\pi^-$ decay ($\acp{dir}=0$), we can derive a measurement of time-integrated CP violation in $D^0$ mixing ($\acp{ind}$) since the measured time-integrated asymmetry can be approximately expressed as
\begin{equation}\label{acp}
\Acp(D^0\to f) \approx \acp{dir}(D^0\to f) +\frac{\langle t\rangle}{\tau}\ \acp{ind},
\end{equation}
where $f$ indicates a generic final state and $\langle t\rangle/\tau\approx2.28$ is the observed average $D^0$ decay time of the sample in units of $D^0$ lifetimes. We then find
$$\acp{ind} = \bigl(-0.02\pm 0.25\stat\pm0.24\syst\bigr)\%,$$
in agreement with our previous determination of this quantity \cite{paper}.

\subsection{Difference of time-integrated asymmetries in $D^0\to K^+K^-$ and $D^0\to\pi^+\pi^-$ decays}
Building upon the techniques developed for the previous measurement of individual asymmetries in $D^0\to h^+h^-$ ($h=\pi$ or $K$) decays \cite{paper}, CDF updated and optimized the analysis toward the measurement of the difference of asymmetries, $\Delta\Acp = \Acp(D^0\to K^+K^-) - \Acp(\pi^+\pi^-)$. The offline selection has been loosened with respect to the measurement of individual asymmetries, since their difference is much less sensitive to instrumental effects allowing for a more inclusive selection, and we now use the full CDF Run~II data sample, which corresponds to $9.7$\,fb$^{-1}$ of integrated luminosity. Requirements on the minimum number of hits for reconstructing tracks are loosened, the $p_T$ threshold for $D$ decay products is lowered from $2.2$ to $2.0$\,GeV/$c$ and $\sim12\%$ fraction of charmed mesons produced in $B$ decays, whose presence does not bias the difference of asymmetries, is now used in the analysis. As a result of the improved selection, the $D^0$ yield nearly doubles and the expected resolution on $\Delta\Acp$ becomes competitive with LHCb's \cite{lhcb}. In the following we briefly present the result, more details can be found in Ref.~\cite{cdf10784}.

\begin{figure}[p]
\centering
\includegraphics[width=0.75\textwidth]{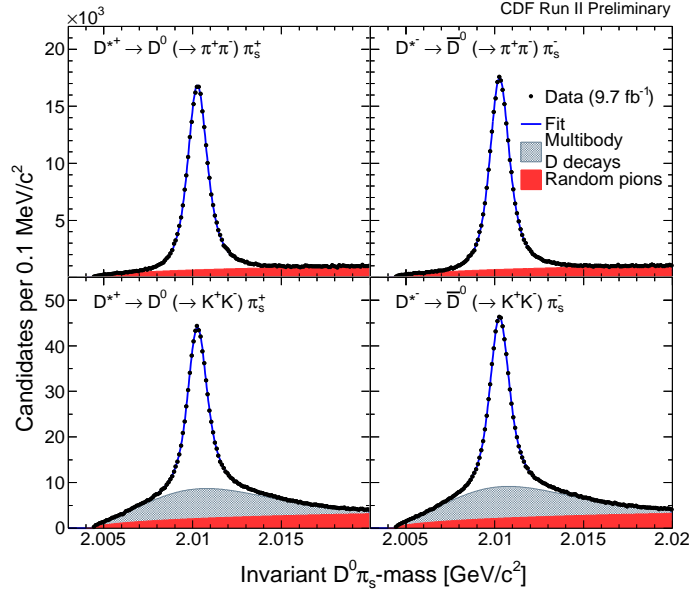}
\caption{Invariant $D^0\pi_s$ mass distribution of $D^*$-tagged $D^0\to \pi^+\pi^-$ (top) and $D^0\to K^+K^-$ (bottom) decays with fit projections overlayed. $D^{*+}\to D^0\pi_s^+$ candidates are on the left, $D^{*-}\to \Dbar^0\pi_s^-$ ones on the right.}\label{fig02}
\end{figure}

The production flavor of the neutral $D$ meson is tagged by the charge of the pion from the $D^{*+}\to D^0\pi^+$ decay. The presence of such an additional ``soft'' (low-momentum) pion causes a bias in the measurement of the asymmetry, induced by a few percent difference in reconstruction efficiency between positive and negative pions at low momentum. However, provided that the relevant kinematic distributions are equalized in the two decay channels, this spurious asymmetry cancels to an excellent level of accuracy in $\Delta\Acp$, leading to systematic uncertainties at the $0.1\%$ level. Using the approximately $550\,000$ $D^*$-tagged $D^0\to\pi^+\pi^-$ and $1.21\cdot10^6$ $D^*$-tagged $D^0\to K^+K^-$ decays shown in Fig.~\ref{fig02}, we measure
$$\Delta\Acp = \bigl(-0.62\pm0.21\stat\pm0.10\syst\bigr)\%,$$
which is $2.7\sigma$ different from zero and consistent with the LHCb result \cite{lhcb}, suggesting that CDF data support CP violation in charm.

\begin{figure}[p]
\centering
\includegraphics[width=0.6\textwidth]{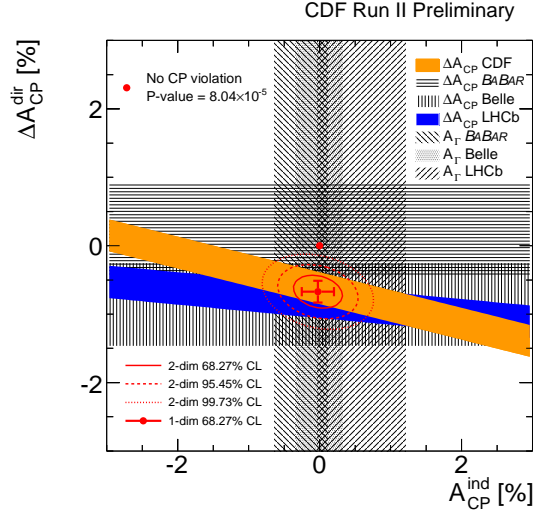}
\caption{Representation of the current knowledge on CP violation in $D^0\to h^+h^-$ decays in the plane ($\acp{ind}$, $\Delta\acp{dir}$). The combination of all results (listed in Ref.~\cite{hfag}) assumes Gaussian, fully uncorrelated uncertainties.}\label{fig03}
\end{figure}

By means of Eq.~\ref{acp} and using the observed values of $\langle t\rangle/\tau\approx2.4$ ($2.65$) for $D^0\to\pi^+\pi^-$ ($D^0\to K^+K^-$) candidates, the observed asymmetry can be combined with all other available measurements of CP violation in $D^0\to h^+h^-$ decays to extract the values of $\acp{ind}$ and $\Delta\acp{dir}=\acp{dir}(D^0\to K^+K^-)-\acp{dir}(D^0\to \pi^+\pi^-)$. The combination, shown graphically in Fig.~\ref{fig03}, yields $\Delta\acp{dir} = (-0.67\pm0.16)\%$ and $\acp{ind} = (-0.02\pm0.22)\%$, which deviates by approximately $3.8\sigma$ from the no-CP violation point.

\end{document}